\documentstyle[12pt]{article}
\def \MSbar {\vbox{\hrule\kern 1pt\hbox{\rm MS}}}
\def \sMSbar {\vbox{\hrule\kern 1pt\hbox{\smallrm MS}}}
\def \GeV { {\ \rm GeV} }

\input epsf.tex
\def\DESepsf(#1 width #2){\epsfxsize=#2 \epsfbox{#1}}
\begin{document}

\title{JETS AND PARTONS}
\author{Davison E.\ Soper\\
{\it Institute of Theoretical Science, University of Oregon}\\
{\it Eugene, Oregon 97403, USA}}

\date{October 18, 1996}

\maketitle

\begin{abstract}
Jet production in high energy hadron-hadron collisions can serve as a probe for
new physics. I review recent data from CDF and D0 on the high $E_T$ jet cross
section. Reporting on recent work of the CTEQ collaboration, I argue that the
apparent excess seen in the CDF data may be due to the gluon distribution
function used in the theoretical calculation being too small at large $x$. I
discuss data on the dijet angular distribution, which shows no sign of a new
physics signal.
\end{abstract}

\vfil
\centerline{\it Adapted from talks at}
\centerline{\it XXVIII International Conference on High Energy Physics,
Warsaw, July 1996}
\centerline{\it and at }
\centerline{\it QCD Euroconference 96, Montpellier, July 1996}

\newpage

%%%%%%%%%%%%%%% BEGIN TEXT %%%%%%%%%%%%%%%%%%%%

\begin{figure}[htb]
\centerline{ \DESepsf(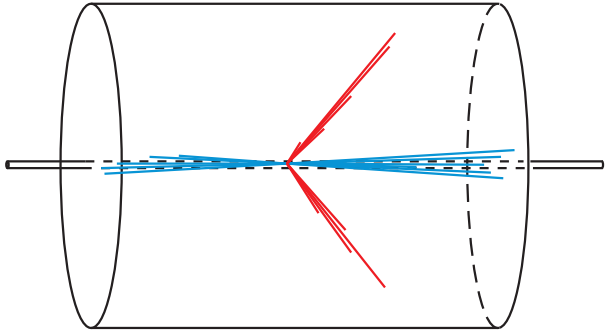 width 8 cm) }
\caption{Sketch of two jet production in a hadron-hadron collision.}
\label{twojets}
\end{figure}

In high energy collisions of hadrons one can produce highly collimated jets
of particles, with the total transverse momentum of a jet reaching several
hundred GeV, as illustrated in Fig.~\ref{twojets}. Such jets reflect the
underlying parton dynamics. 

One can use measurements of jet cross sections to check in detail how well the
perturbative diagrams of QCD describe actual quark and gluon collisions. One
can also use the measurements to help determine parton distributions. In
particular, there is an opportunity to pin down the comparatively unknown the
gluon distribution, since gluon initiated processes are important for moderate
values of the jet transverse momentum. As we will see, the data may also be
telling us something surprising about the gluon distribution at large $x$.

\begin{figure}[htb]
\centerline{ \DESepsf(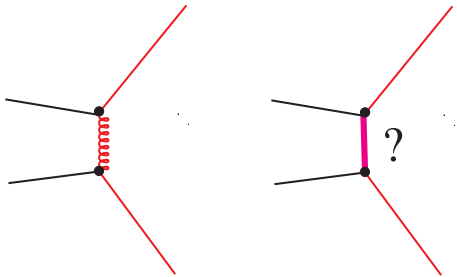 width 6 cm) }
\caption{Quark-quark scattering by gluon exchange and by a hypothetical new
interaction.}
\label{newphys}
\end{figure}

The most important role that can be played by jet cross sections is to help
us to discover a breakdown of the Standard Model at small distances. Suppose
that there is some new process that can lead to parton-parton scattering, as
illustrated in Fig.~\ref{newphys}. Such a new process could be part of a
theory that has the Standard Model as its long distance limit. Then the new
physics signals accessible at moderately long distances can be characterized
as additions $\Delta{\cal L}$ to the lagrangian of the Standard Model, for
example
\begin{equation}
{{\Delta{\cal L} = {\tilde g^2 \over \Lambda^2} \
\bar \psi \gamma^\mu \psi\
\bar \psi \gamma_\mu \psi.}}
\label{calL}
\end{equation}
In contrast to the usual dimension 4 terms in the lagrangian, such a term has
dimension 6 and thus a factor of $1/\Lambda^2$, where $\Lambda$ is a large mass
characteristic of the short distance scale. Here $\tilde g^2$ is a coupling of
the new theory.

The effect of such a term $\Delta{\cal L}$ in the lagrangian is to change the
inclusive cross section $d\sigma/d E_T$ to make a jet with transverse energy
$E_T$. One looks for a signal of the form 
\begin{equation}
{{{{\rm Data} - {\rm Theory} \over {\rm Theory}}}}
\propto \tilde g^2 { {{E_T^2}} \over \Lambda^2}.
\end{equation}
The ``Theory'' here is next-to-leading order QCD~\cite{EKS,ACGG,GGK}, not
including a new physics contribution.

Jet cross sections at hadron colliders are a good place to look for traces of
$\Delta{\cal L}$ because one can go out to very large $E_T$, over 400 GeV at
the Fermilab Collider, and still have a measurable Standard Model signal. The
one jet inclusive cross section is useful because it is the most inclusive jet
cross section, but (as will become clear in this talk) the two jet angular
distribution is also very useful.

In order to have reasonably precise Standard Model predictions, one needs to be
careful to state the algorithm by which jets are defined. The choice of
algorithm is a matter of convention, but the convention must be the same in
the theory and in the experiment. In hadron-hadron physics, the definition
that has been typically used is the ``Snowmass Convention''~\cite{snowmass},
in which a jet consists of all the particles inside a certain cone.

In the theoretical calculation, the jet cross section takes the form
\begin{eqnarray}
\lefteqn{
{d\sigma \over  d E_T} =
\int d\xi_a\ {{f_{a/A}(\xi_a)}}\ 
 \int d\xi_b\ {{f_{b/B}(\xi_b)}} 
}
\nonumber\\
\lefteqn{ \times \Biggl\{
 \int dk_1 dk_2 {d \sigma(a+b \to 1 + 2) \over d k_1 \ d k_2}\
{{{\cal S}_2}}(k_1,k_2; E_T)}
\nonumber\\
&& +
 \int dk_1 dk_2 dk_3{d \sigma(a+b \to 1 + 2 + 3) \over
 d k_1 \ d k_2 \ d k_3}\
%\nonumber\\
%&&\hskip 2 cm \times
{{{\cal S}_3}}(k_1,k_2,k_3; E_T)
\Biggr\}
\end{eqnarray}
Here ${f_{a/A}(\xi)}$ is the parton distribution function giving the
probability to find a parton of type $a$ in a hadron of type $A$ carrying
momentum fraction $\xi$. The function ${d \sigma(a+b \to 1 + \cdots + N) / [d
k_1 \cdots d k_N}]$ gives the QCD cross section for partons $a$ and $b$ to make
$N$ final state partons. Finally the functions ${{{\cal S}_N}}$ contain the
algorithm for defining the jet cross section. The calculation will produce a
finite answer because the functions ${{{\cal S}_N}}$ are ``infrared safe''.
{\it Eg.}
\begin{equation}
{{{\cal S}_3}}(k_1^\mu,(1-\lambda)k_2^\mu,\lambda k_2^\mu; E_T)
=
{{{\cal S}_2}}(k_1^\mu,k_2^\mu; E_T)
\end{equation}
This says that a mother parton will produce the same contribution to the jet
cross section whether or not it breaks into two daughter partons moving in
the same direction.

The theoretical calculation is strictly an order $\alpha_s^3$ calculation.
Monte Carlo event generators can do a good job of modeling the part of
the order $\alpha_s^4$, $\alpha_s^5$, $\alpha_s^6$, \dots contributions that
comes from approximately collinear parton branchings, but because of the
property of infrared safety enjoyed by the jet cross section, the approximately
collinear integration regions are not particularly important. For this
reason, no $\alpha_s^4$, $\alpha_s^5$, $\alpha_s^6$, \dots contributions are
included. Also because of infrared safety, jet cross sections at high $E_T$
are {{not}} sensitive to hadronization. Thus the calculations do not try to
correct for hadronization.

\begin{figure}[htb]
\centerline{ \DESepsf(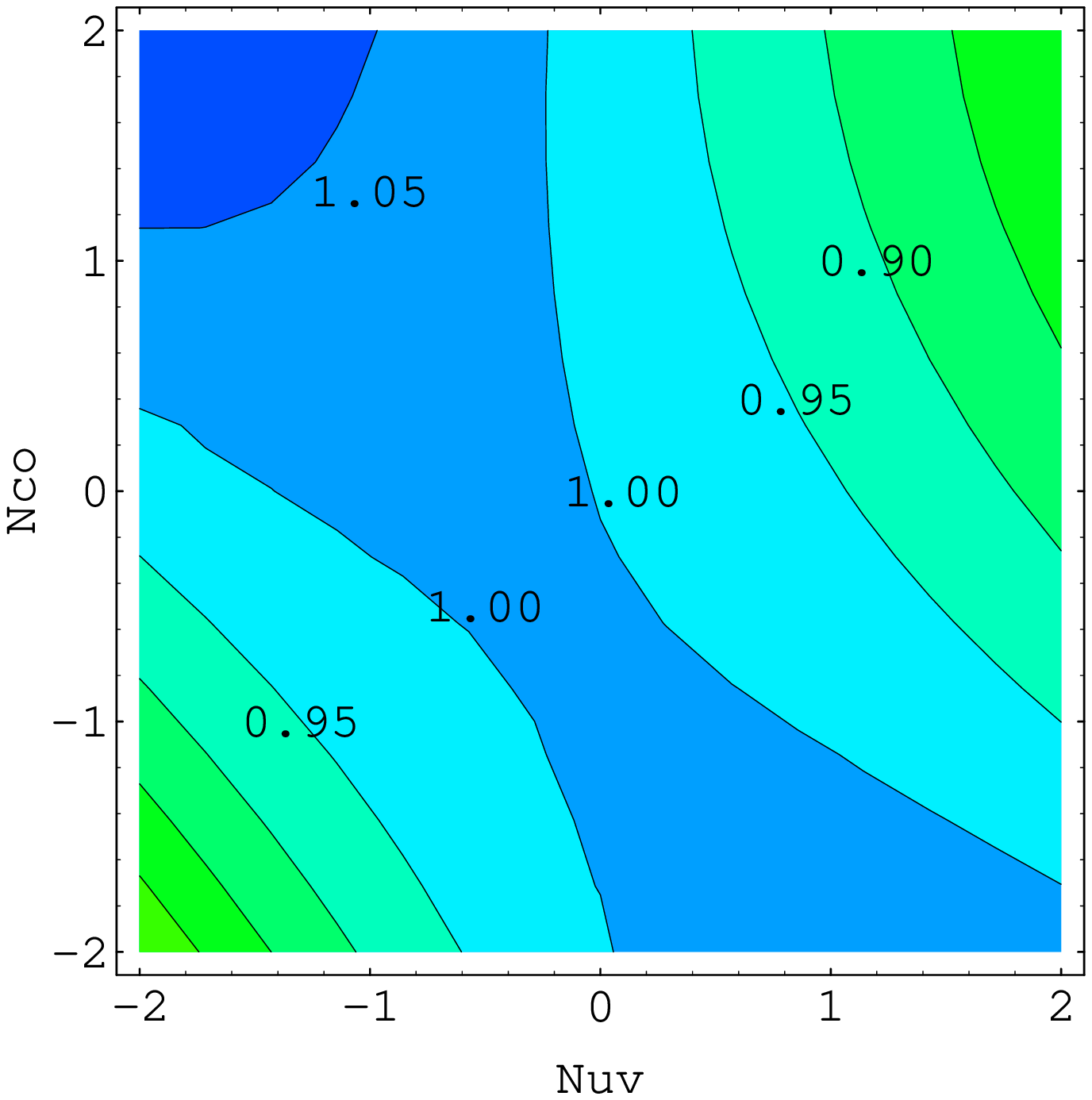 width 6.5 cm)\DESepsf(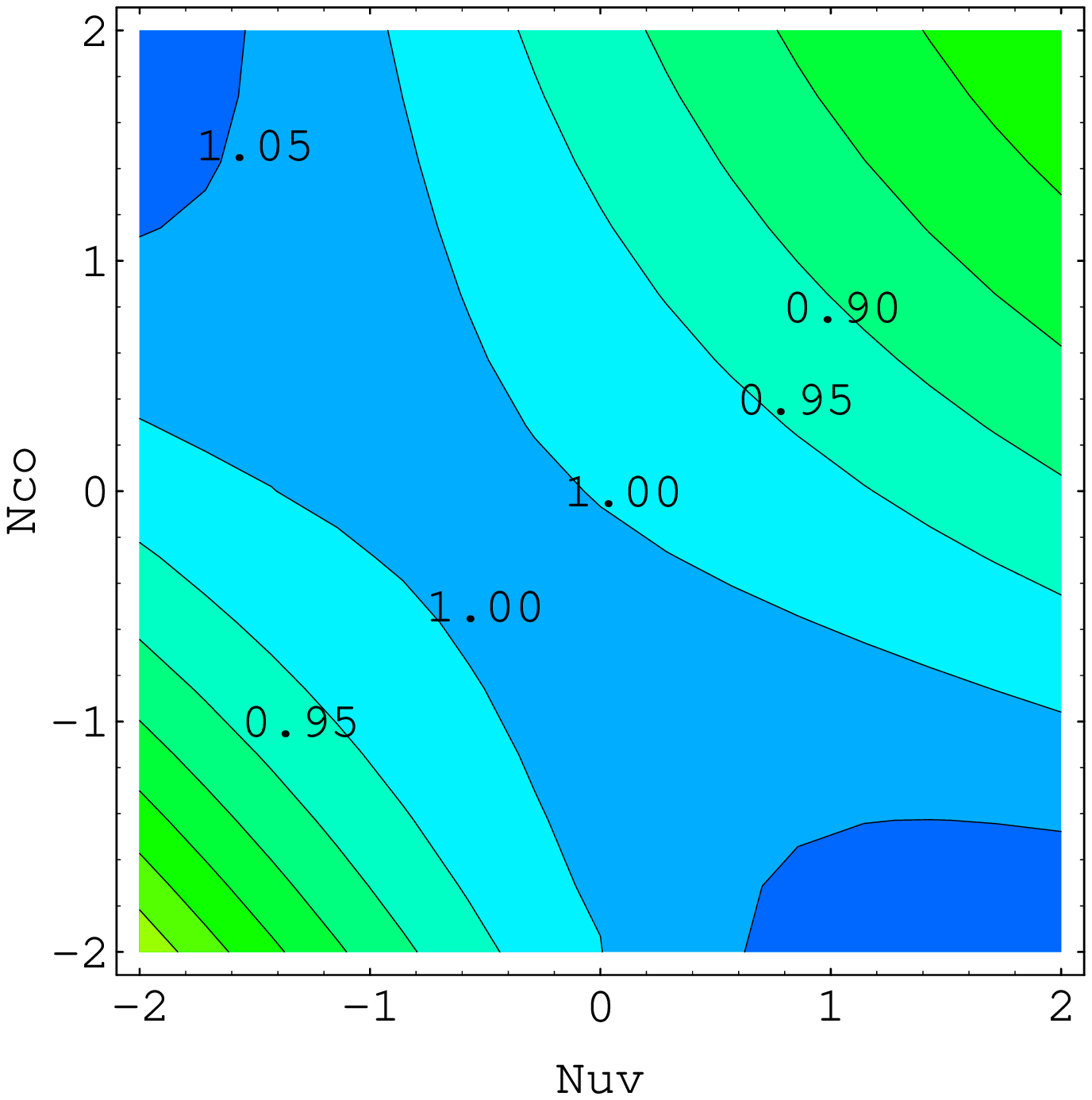 width 6.5 cm) }
\caption{Scale dependence for $E_T = 100\ {\rm GeV}$ (left) and
 $E_T = 500\ {\rm GeV}$ (right).}
\label{fig:mu100}
\end{figure}

The theoretical formula for the one jet inclusive cross section depends on a
scale $\mu_{\rm UV}$ that appears in the running coupling and on another scale
$\mu_{\rm CO}$ that appears in the parton distribution functions. These scales
also appear explicitly in the order $\alpha_s^3$ contributions to the cross
section. In Fig.~\ref{fig:mu100}, I show the dependence
of the calculated cross section on the logarithms of the
scales $\mu$, $N_{\rm UV}$ and $N_{\rm CO}$ defined by $\mu_{\rm UV}=
(E_T/2)\times 2^{N_{\rm UV}}$, $\mu_{\rm CO}= (E_T/2)\times 2^{N_{\rm CO}}$.
The figures are contour graphs with 5\% contour lines of $d\sigma/dE_T\,d\eta$
with arbitrary normalization, with the jet rapidity $\eta$ set equal to zero.
We see that both at $E_T = 100\ {\rm GeV}$ and at $E_T = 500\ {\rm GeV}$ the
scale dependence is on the order of 15\%. This provides a rough estimate of the
likely error in the theory due to leaving out the uncalculated contributions of
order  $\alpha_s^4$ and higher.

\begin{figure}[hbt]
\centerline{ \DESepsf(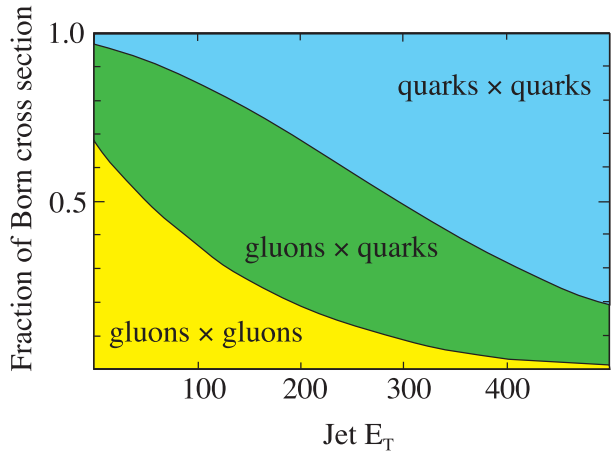 width 8.0 cm) }
\caption{Contribution to the Born-level one jet inclusive cross section
from gluon-gluon, gluon-quark, and quark-quark collisions.}
\label{fig:gluefraction}
\end{figure}

The jet cross section at moderate $E_T$ is sensitive to the gluon
distribution, while at large $E_T$, it is mostly the well measured quark
distributions that count. This is illustrated in Fig.~\ref{fig:gluefraction}.
Notice, however, that even at the highest values of $E_T$, the contribution
from gluons is still not negligible.

\begin{figure}[p]
\centerline{ \DESepsf(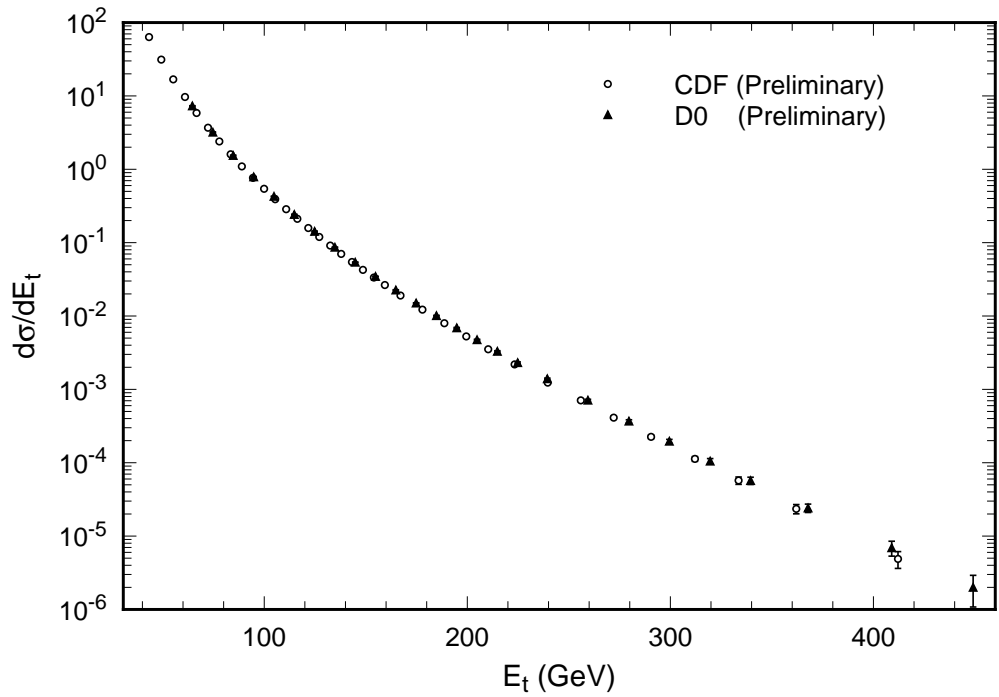 width 14.0 cm) }
\caption{CDF and D0 cross sections $\langle d\sigma/ d\eta\, dE_T\rangle$
averaged over $0.1 < |\eta| < 0.7$ (CDF) and $ |\eta| < 0.5$.}
\label{fig:JetData}
\end{figure}

Let us now look at the data for the one jet inclusive cross section. In
Fig.~\ref{fig:JetData}, I show the CDF~\cite{CDF} and D0~\cite{D0} cross
sections as a function of $E_T$. The range of rapidities included in the
two cases is not same, so that the data are not strictly comparable.
Nevertheless, one has the impression of good agreement between the two
experiments.

\begin{figure}[htb]
\centerline{ \DESepsf(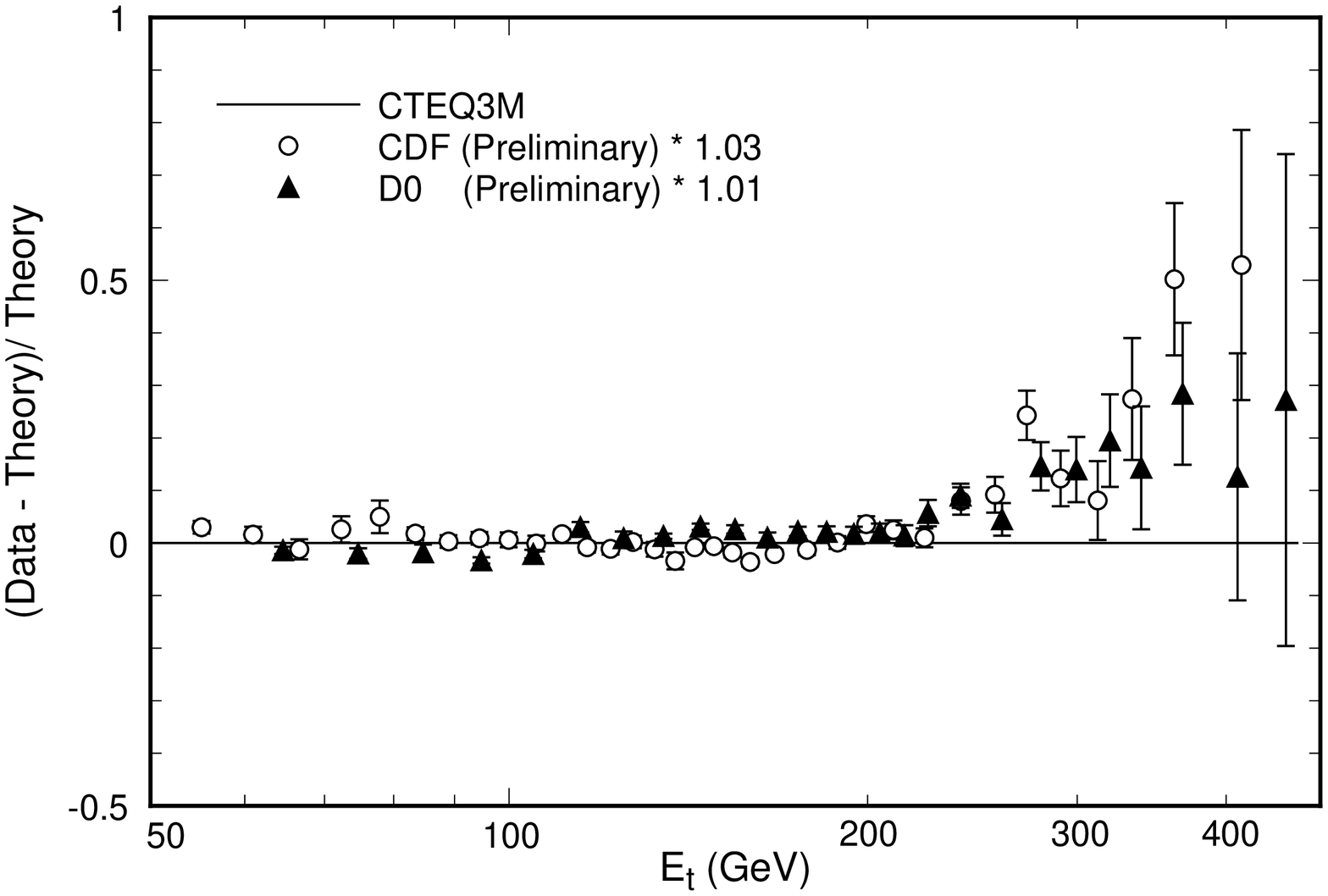 width 9.0 cm) }
\caption{Comparison~\protect\cite{CTEQ4} of CDF and D0 one jet inclusive
cross sections with theory using CTEQ3M partons.}
\label{fig:JetcM}
\end{figure}

In Fig.~\ref{fig:JetcM}, I show a comparison of these data to theory, taken
from the work of the CTEQ Collaboration~\cite{CTEQjet,CTEQ4}. The comparison
uses the CTEQ3M set of parton distributions. The (Data $-$ Theory) /Theory
format of the figure allows us to see the quality of the agreement despite
the fact that the data falls by seven orders of magnitude in the $E_T$ range
shown. The systematic experimental errors are not shown. In the CDF data, the
systematic error is about 20\%. For the D0 data, the systematic error is
larger.

\begin{figure}[hbt]
\centerline{ \DESepsf(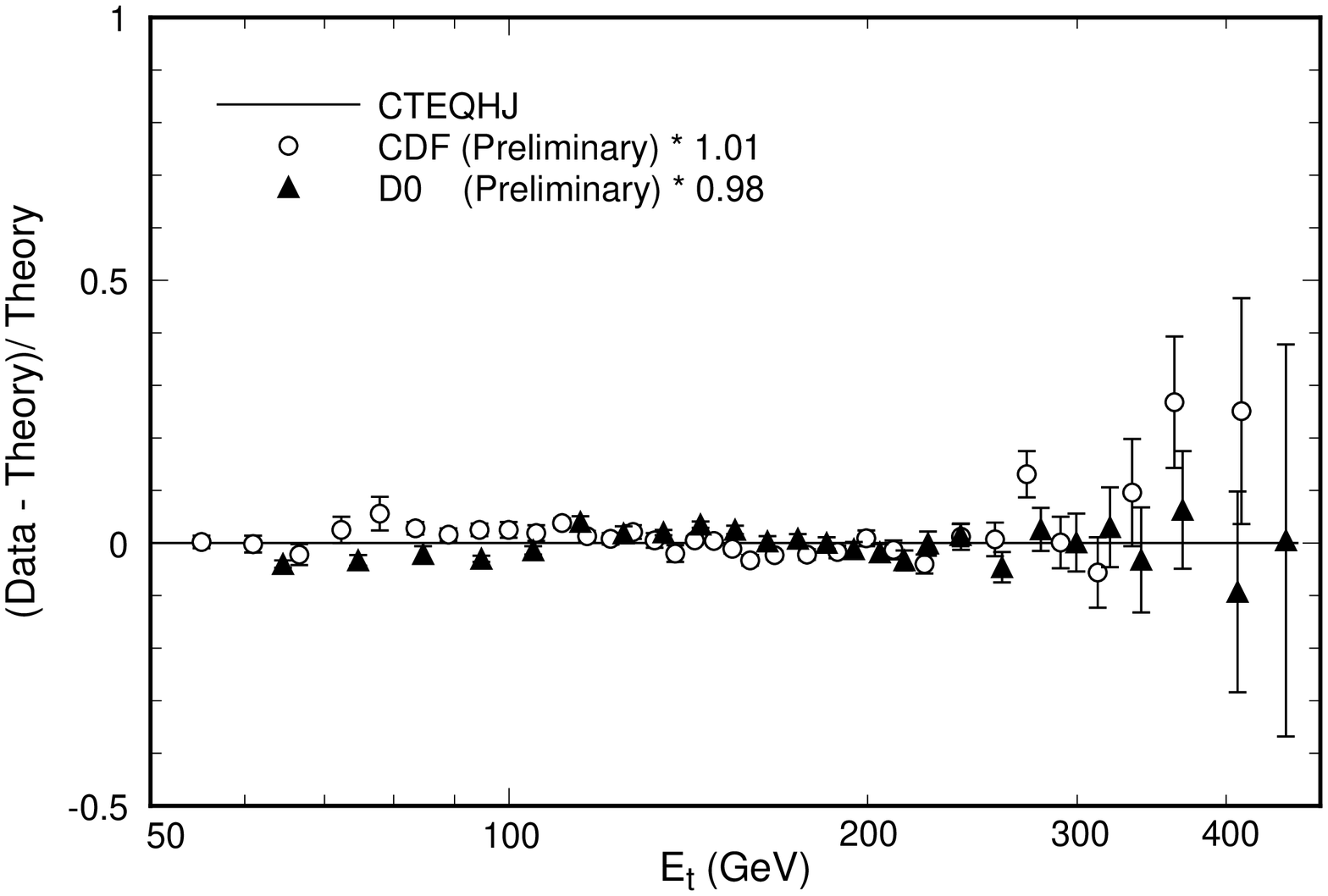 width 9.0 cm) }
\caption{Comparison~\protect\cite{CTEQ4} of CDF and D0 one jet inclusive cross
sections with theory using specially fit CTEQ4HJ partons.}
\label{fig:JetHJ}
\end{figure}

We see from the graph that there is very good agreement between theory and
experiment for $50 \GeV < E_T < 250 \GeV$. For $250 \GeV < E_T$, the D0
data is in agreement with the CDF data within the errors. Furthermore, the D0
data is in agreement with the theory within the errors. However the more
precise CDF data shows a systematic rise above the theory as $E_T$ increases.
Using an earlier data set than shown above and comparing to MRSD0$^\prime$
partons, the CDF collaboration reports~\cite{CDF} that the chance that the
high $E_T$ excess is a statistical fluctuation, taking the systematic error
into account, is less than 1\%. This is just the sort of new physics signal
that we were looking for, so it is well worth a careful examination. 

We can also compare to QCD theory using the latest CTEQ4M parton
distributions~\cite{CTEQ4}. This set makes use of new data on
deeply inelastic scattering from NMC, E665, Zeus and H1. The data included in
the fit also includes the CDF and D0 jet data under discussion. The jet data,
however, do not have enough statistical power to control the fit. The
resulting (Data $-$ Theory) / Theory plot looks much the same as
Fig.~\ref{fig:JetcM}. One can do the same using the latest MRS parton
distribution set~\cite{MRS}, which uses the new deeply inelastic
scattering data but not the jet data. The result is the same.

It seems that the theoretical prediction is robust against changes in the
parton distribution, but we may ask more directly whether there could be
enough flexibility in the partons to account for the apparent high $E_T$
excess. Note that high $E_T$ corresponds to high momentum fraction $x$ for
the colliding partons. $E_T \approx 450 \GeV$ corresponds to $x \approx 0.5$.
The quark distribution at large $x$ is quite accurately pinned down by deeply
inelastic scattering. The gluon distribution is not. Since with ``standard''
partons, events with initial gluons do not account for much of the cross
section, one would have to change the gluon distribution quite drastically at
large $x$ in order to account for the high $E_T$ excess. The CTEQ
Collaboration has tried this~\cite{CTEQjet}.  They forced the parton
parameterization to change so that the theory cross section goes through the
large $E_T$ jet data, while the fit to deeply inelastic scattering and other
data remains as good as possible. The result is shown in Fig.~\ref{fig:JetHJ}.
In order to obtain this result, the quark distributions remained nearly
unchanged, while the gluon distribution function approximately doubled near $x
= 0.5$.

We see that with the modified parton distributions the jet data can be
accommodated. The question is, does this ruin agreement with other data?
The $\chi^2$ for the CTEQ4M set compared to 1297 DIS and Drell-Yan data was
1320. The $\chi^2$ for the special CTEQ4HJ set compared to this same data was
1343. This is not as good, but not really much worse. Since systematic errors
are not generally included in these $\chi^2$ values, the change may be
regarded as not significant.

\begin{figure}[htbp]
\centerline{ \DESepsf(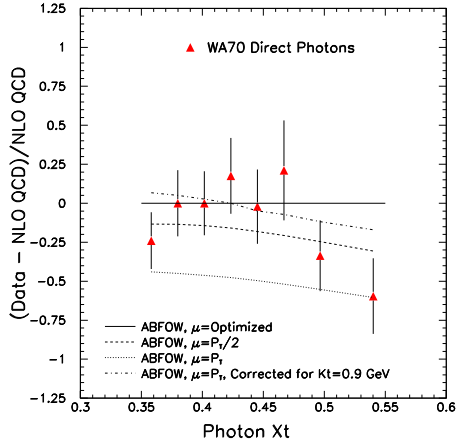 width 7.0 cm) }
\caption{Direct photon production. Data from the WA70 experiment is 
compared~\protect\cite{CTEQjet} to theory using ABFOW partons, taking the scale
$\mu$ in $\alpha_s$ and the parton distributions to be fixed by the ``principle of minimal sensitivity.''
Three other theory curves show the effect of choosing $\mu = P_T$ and $\mu =
P_T/2$ instead and of choosing $\mu = P_T$ while adding smearing in the
transverse momentum for the incoming partons.}
\label{fig:WA70kt}
\end{figure}

\begin{figure}[htbp]
\centerline{ \DESepsf(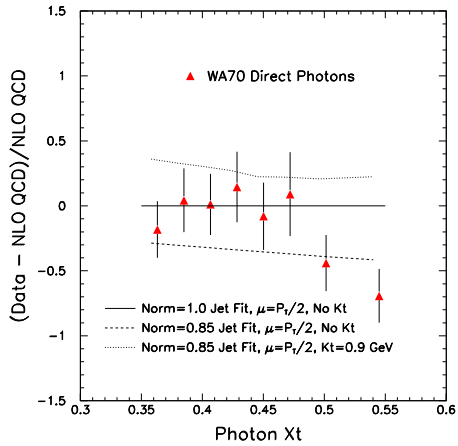 width 7.0 cm) }
\caption{Direct photon production. Data from the WA70 experiment is 
compared~\protect\cite{CTEQjet} to theory using CTEQ4HJ partons (the ``Norm =
1.0 Jet Fit'') and also to the theory using an alternative set of partons that
give a good fit to the high $E_T$ jet data (the ``Norm = 0.85 Jet Fit'').
The scale $\mu$ is set to $P_T/2$, and a theory curve using transverse momentum
smearing is also shown.}
\label{fig:WA70HJ}
\end{figure}

What about direct photon production in hadron-hadron collisions? This process
gets big contributions from $quark + gluon \to photon + X$, so it is sensitive
to the gluon distribution. Furthermore, experiments at fixed target energies
can reach to $x_\gamma \equiv 2 P_T/\sqrt s$ on the order of 1/2, which probes
the gluon distribution with $x \approx 1/2$.  In Fig.~\ref{fig:WA70kt}, I show
the comparison of theory and experiment for WA70 data~\cite{WA70} using
the conventional ABFOW parton distribution set~\cite{ABFOW}. One sees that the
agreement between theory and experiment is none too good. The figure also
shows three alternative theory curves based on different scale choices and on
the addition of transverse momentum effects that might be expected from
multiple gluon emission in the initial state. One sees that in fact the
theory is quite unstable. This instability may be attributed to the fact that
the transverse momenta involved ($< 7 \GeV$) are not large. In
Fig.~\ref{fig:WA70HJ}, I show the WA70 data compared to a theoretical
calculation using the CTEQ4HJ set. Again, the agreement between theory and
experiment is none too good. But it is not much worse than with conventional
partons, and one can argue that the agreement is good enough given the
theoretical uncertainties.

I conclude that direct photon production results do not, in fact, determine
the gluon distribution very well.

Let me mention two issues raised by recent theoretical papers. First, Klasen
and Kramer~\cite{KK} have investigated how the prediction for the high $E_T$
jet cross section depends on whether the fitting of partons and then the
subsequent calculation of the jet cross section is done in the \MSbar\
factorization scheme or in the DIS scheme. They find that the difference is
large and that the high $E_T$ excess goes away when the DIS scheme is used.
My interpretation of this is that the difference comes in the parton fitting.
Since the gluon distribution at large $x$ is poorly constrained by data, the
results of the fitting program can be different depending on small
differences in the way the fitting is done. Second, in the large $x$ region,
where the parton distribution functions are steeply falling, fixed order
perturbation theory may be inadequate and one may need a soft gluon
summation. This has not yet been done for jet production, but it has been
done for top quark production by two groups~\cite{catani,berger}. The effect
appears to be moderately large~\cite{berger} or small~\cite{catani} depending
on the method of calculation.

We have seen that there is an excess of high $E_T$ jets compared to standard
theory, but that this excess has a plausible explanation based on the standard
gluon distribution function being too small at large $x$. Fortunately, there
is a way to get at the possible new physics signal that is not very sensitive
to the parton distributions. One can look at the jet production and examine the
angular distribution of the two jets in each event that have the highest
$E_T$s. Specifically, consider the cross section
\begin{equation}
{d\,\sigma \over d\,M_{JJ}\ d\,\eta _{JJ}\ d\,{{\eta^*}}} \ .
\end{equation}
Here $M_{JJ}$ is the jet-jet mass, $\eta _{JJ} = (\eta_1 + \eta_2)/2$ is the
rapidity of the jet-jet c.m. system,  and ${{{\eta ^*} = (\eta_1 -
\eta_2)/2}}$ is the rapidity ($ - \ln\tan(\Theta^*/2)$) of first jet as viewed
in the jet-jet c.m. system.   Look at the cross section as a function of ${{
\eta^*}} $  for a fixed bin of $ M_{JJ} $ and $\eta _{JJ}$. Dividing by the
cross section integrated over $\eta*$ gives the angular distribution. Since
the angular distributions in quark-quark collisions, quark-gluon collisions,
and gluon-gluon collisions are very similar, the net angular distribution is
not very sensitive to the parton distribution functions.

Fortunately, the angular distribution {\it is} sensitive to a new physics
signal associated with a term $\Delta{\cal L}$, such as that in
Eq.~\ref{calL}. Vector boson exchange in QCD produces an angular distribution
with the behavior
\begin{equation}
{d\,\sigma \over \ d\,\eta^*} \propto \cosh(2 \eta^*) 
\hskip 1.0 cm
\eta^* \gg 1 \ .
\end{equation}
A new physics term gives low angular momentum partial waves and thus
few events with $ \eta^* >1$.

\begin{figure}[htbp]
\centerline{ \DESepsf(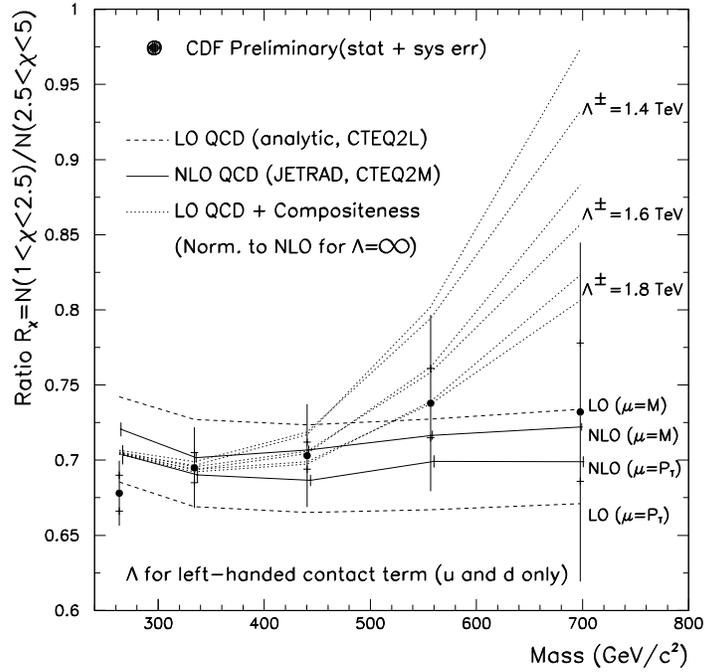 width 9.0 cm) }
\caption{ Dijet angular ratio as a function of the jet-jet mass. Theory
curves show QCD predictions at next-to-leading order (NLO) and also at leading
order. Also shown are expectations from including a new physics term with
various values for the dimensionful parameter giving its strength.
This plot is an earlier version of the CDF
plot~\protect\cite{CDF2jet} submitted for publication.}
\label{fig:jetangle}
\end{figure}

To look at this question, the CDF group~\cite{CDF2jet} has studied the ratio
\begin{equation}
{{{\cal R}}}=
\int_{{0}}^{{{0.46}}}\!\!\!d\eta^*\ 
{d\,\sigma \over d\,M_{JJ}\ d\,\eta^*} \ \biggl/
\int_{{{0.46}}}^{{{0.80}}}\!\!\!d\eta^*\ 
{d\,\sigma \over d\,M_{JJ}\ d\,\eta^*} .
\end{equation}
In Fig.~\ref{fig:jetangle}, I show the CDF result. The data are compared to
standard QCD theory (with standard partons) and to standard QCD plus a new
physics term. The new physics term is similar to that in Eq.~(\ref{calL}). Its
strength is parameterized by a parameter $\Lambda^+$, which is essentially  the
$\Lambda$ in Eq.~(\ref{calL}) with a conventional choice for $\tilde g$ if we
choose the sign so that the new physics term interferes constructively with
standard QCD. If we choose destructive interference, the new physics signal is
parameterized by a parameter $\Lambda^-$. Choosing for the moment positive
interference,  the CDF single jet inclusive cross section favors $\Lambda^+
\approx $ 1.6 TeV. The angular distribution data rule out this value of
$\Lambda^+$ at the 95\% confidence level~\cite{CDF2jet}. A somewhat larger
value of $\Lambda^+$ could still be consistent with both sets of data and QCD
plus new physics with standard parton distributions. Choosing destructive
interference, the conflict between the angular distribution data and the
single jet data, taking standard partons in the theory, is not as strong.
I have discussed the CDF data here, but D0 data~\cite{D0dijet} on the dijet
angular distribution also shows no sign of a new physics signal.
 
I conclude that the angular distribution data disfavor the new physics
hypothesis as the explanation of the high $E_T$ excess seen in the single jet
cross section. The ``more gluons'' hypothesis remains as a plausible
explanation. It also seems plausible that part of the explanation lies in
corrections to the theory from soft gluon summation.

\section*{Acknowledgements}

This work was supported in part by U.S.\ Department of Energy grant
DE-FG03-96ER40969.

%***************************

%***************************
%***************************

\end{document}